# Measuring the Adhesion of Graphene Flake Networks via Button Shear Tests

*Jorge Eduardo Adatti Estévez, Josef Schätz, Jasper Ruhkopf, Annika Weber, David Tumpold, Alexander Zöpfl, Ulrich Krumbein, Max Christian Lemme\**

*J. E. Adatti Estévez, A. Weber, D. Tumpold, U. Krumbein*
*Infineon Technologies AG*
*Am Campeon 6, 85579 Neubiberg, Germany*

*J. E. Adatti Estévez, J. Schätz, J. Ruhkopf, M. C. Lemme*
*Chair of Electronic Devices, RWTH Aachen University*
*Otto-Blumenthal-Straße 25, 52074 Aachen, Germany*

*J. Schätz, A. Zöpfl*
*Infineon Technologies AG*
*Wernerwerkstraße 2, 93049 Regensburg, Germany*

*J. Ruhkopf, M. Lemme*
*AMO GmbH*
*Otto-Blumenthal-Straße 25, 52074 Aachen, Germany*

*\*email: max.lemme@eld.rwth-aachen.de*

**Abstract**

Graphene flake-based dispersions are attractive materials for various applications in microelectronics because of their ease of fabrication and the potential to deposit them on diverse substrates. The integration of these materials into conductive networks and microdevices requires thorough knowledge of their mechanical material properties, including adhesion. This paper presents quantitative adhesion measurements of graphene flake networks on silicon dioxide ($SiO_2$) via button shear testing (BST). In this method, shear forces are applied to prefabricated micrometric buttons until they delaminate, providing information about the shear strength of the underlying graphene. We applied BST to graphene flake networks with different flake structures and defect densities. Flat flakes, a flat network structure, and a high flake defect density improve adhesion. We further demonstrate that graphene flake networks have stronger adhesion than chemical vapor deposited (CVD) monolayer graphene grown on copper and transferred to $SiO_2$. Hexamethyldisilazane (HMDS) increases the total adhesion force by improving flake network formation. Finally, we provide flake-type-specific delamination patterns by combining BST, optical microscopy, and Raman spectroscopy. We establish BST as a quantitative technique for measuring the adhesion of graphene dispersions and show the crucial role of interflake junctions in the overall adhesion of graphene flake networks.

# 1. Introduction

Graphene-based electronic devices have been proposed for a wide range of sensor applications because of their mechanical and electronic properties[1–3]. A significant challenge toward their commercialization lies in the integration of large-area graphene films into standard semiconductor technology[4–6], particularly wafer-level growth and transfer[7,8]. Graphene flakes incorporated in dispersions present a scalable and low-cost alternative[9–12] that can be easily deposited on silicon wafers and many other substrates[13,14] without the need for complex transfer techniques. In fact, graphene dispersions have been proposed for devices such as gas sensors[15], humidity sensors[16], electrochemical biosensors[17], and energy storage devices[18,19].

Graphene dispersions can be deposited to form conductive graphene flake networks via simple methods such as inkjet printing[20,21], spray coating[22–24], or spin coating[25–27]. In addition, graphene flakes differ in structure and defect density depending on their fabrication method[11,12,28–30], which allows tuning dispersions according to their targeted application. Therefore, the selection of graphene flakes requires thorough characterization.

Adhesion is a crucial factor when graphene is integrated into semiconductor devices[31–33], with direct implications for device reliability[34–37]. It has been studied via various methods, including blister tests[38,39], nanoparticle intercalation[34], microforce sensor-assisted shearing[40], nano-scratch testing[41], and atomic force microscopy (AFM)-based techniques such as dynamic force mapping[42], nanoindentation[43], and friction measurement[44]. However, these methods are time-consuming and/or not suitable for adhesion measurements of graphene flake networks. The adhesion of graphene flake networks has been assessed by pull tests with adhesive tape[37,45,46] or through ultrasonication[45,47], and their electrical resistance has been measured before and after delamination. Although both methods are simple to implement, they remain qualitative.

Button shear testing[48–50] (BST) has recently been shown to be an effective method for measuring the adhesion of two-dimensional (2D) materials[51], such as carbon vapor deposited (CVD) graphene, hexagonal boron nitride (hBN) and molybdenum disulfide ($MoS_2$). BST is a well-established measurement technique for bulk materials that provides simple, quantitative adhesion information in the form of critical shear force values. It involves the application of a lateral shear force on a button until it is detached from the substrate. The force required for detachment serves as a measure of the adhesion between the button material and the substrate. Here, we propose BST for characterizing the adhesion of graphene dispersions. We demonstrate its feasibility with different dispersions and graphene flake types. We show that the graphene flake structure and the defect density are critical to adhesion. The quantitative nature of BST further allows a direct comparison between graphene dispersions and CVD graphene, as well as a validation and assessment of the adhesion-promoting effect of hexamethyldisilazane (HMDS). Finally, we establish that BST generates delamination patterns specific to the graphene flake type. Their evaluation provides insights into the key role of interflake adhesion in overall flake network adhesion.

## 2. Results and Discussion

### 2.1 Button fabrication and button shear testing

All the experiments were conducted on 2 × 8 $cm^2$ silicon chips with a 280 nm thermally grown silicon dioxide layer. We used two dispersions based on organic solvents and one based on water. Each dispersion had a different graphene flake type, whose defect density and structure were characterized via Raman spectroscopy, scanning electron microscopy (SEM), and atomic force microscopy (AFM) (see Section 2.2). The adhesion-promoting effect of HMDS was also investigated by depositing a graphene dispersion on an HMDS-coated chip, as well as a PMMA film without graphene as a reference.

First, the chips were exposed to an oxygen ($O_2$) plasma, followed by spin-coating of the graphene dispersions (**Figure 1a**) and thermal annealing. The samples were then spin-coated with 5 μm polymethyl methacrylate (PMMA), after which a 40 nm thick aluminum (Al) layer was evaporated on top of the PMMA (**Figure 1b**). Rectangular buttons with dimensions of $100 \times 60$ μm$^2$ were defined via optical lithography. An alkaline-based developer was used to develop the resist and to etch the Al in the same fabrication step. Finally, the PMMA and graphene next to the buttons were removed with $O_2$ plasma (**Figure 1c**).

We fabricated a reference sample without graphene (PMMA$_{only}$) as well as a further sample with Med$_{Org}$ using HMDS as an adhesion promoter (Med$_{OrgHMDS}$). The fabrication of the PMMA$_{only}$ reference followed the same procedure explained above, except for the spin-coating of the graphene dispersion. For Med$_{OrgHMDS}$, a thin layer of HMDS was evaporated on the substrate prior to spin-coating the graphene dispersion. The subsequent steps were carried out in the same manner as those for the graphene samples. Further details on the sample fabrication process can be found in the Methods section.

We used a DAGE4000Plus pull-shear tester from Nordson Corporation to perform the button shear tests. For shearing, the shear head is placed in front of the button and lifted above the substrate, as illustrated in **Figure 1d**. The sample stage then starts to move toward the shear head, and both the force acting on the shear head and the displacement of the sample stage are recorded. Details on the measurement parameters can be found in the Methods section. **Figure 1e** and **Figure 1f** are 3D laser scanning micrographs of a button before and after shearing, and the direction of the shear force is indicated in the latter. The micrographs show that the graphene flake network was not completely delaminated, as the region where the button was initially patterned was clearly visible. The detached button also partially covers the remaining graphene because the movement of the shearing head stops shortly after the button is sheared off.

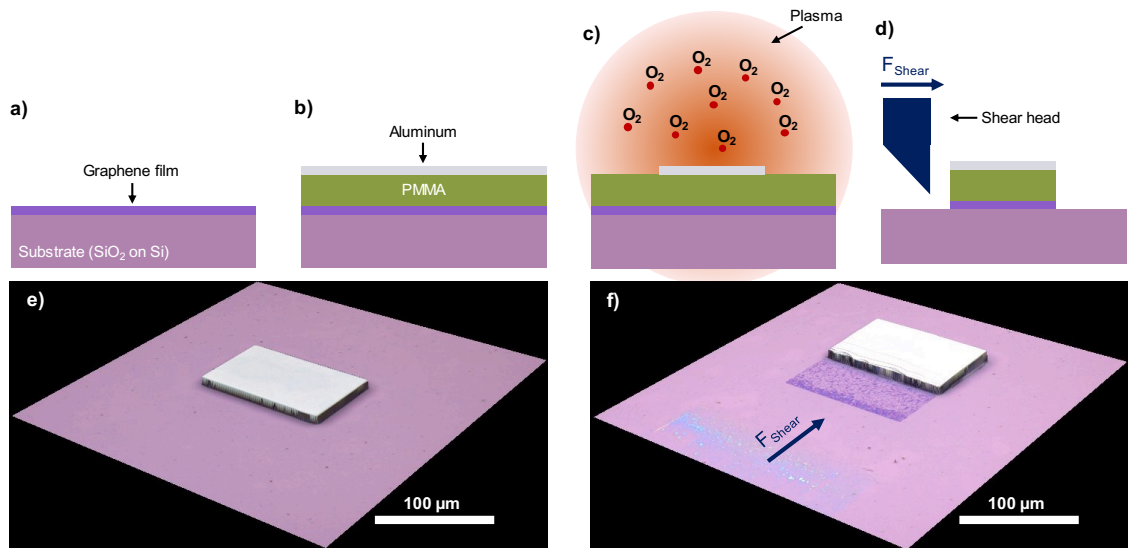

**Figure 1**: Schematic representation of the button fabrication process and 3D micrographs of the shearing process. a) Graphene dispersion is deposited on the substrate via spin-coating. The sample then undergoes thermal annealing, leaving a thin graphene film (< 200 nm) on the substrate. b) A 5 μm-thick PMMA layer is deposited via spin-coating on the graphene surface, and a 40 nm thick Al layer is evaporated on top. c) The optical lithography step enables the definition of the button pattern on the Al layer because of the Al-etching properties of the alkaline developer. Subsequently, $O_2$ plasma removes the PMMA and graphene outside of the button. d) Start of BST: The shear head is placed in front of the button and lifted over the surface. The movement of the button toward the shear head and the recording of force and displacement start. e) 3D micrograph of the button before BST. f) Displaced button and graphene residues after BST; the shearing force is applied to the long button edge.

## 2.2 Characterization of the dispersions used

**Figure 2a** shows the Raman spectra of the three dispersions used. The Raman analysis focuses on the ratio between the D- and G-peak intensities ($I_D/I_G$), as this parameter is correlated with the defect density as well as the oxygen (O) and hydrogen (H) functional group contents[30,52–54]. The three dispersions (and the samples composed of them) were named according to their defect density, which includes edge defects among various other defect types and solvents. Med$_{Org}$ and Low$_{Org}$ are both based on organic solvents, and their flakes exhibited medium and low defect densities, respectively. High$_{H2O}$ consisted of highly defective graphene flakes dispersed in water. The Raman measurements were performed on SiO$_2$-on-Si substrates with drop-cast graphene dispersions. After dispersion deposition, the samples were thermally annealed according to the parameters in the Methods section.

The flake structures were visualized via SEM, whereas AFM was used to measure the surface roughness of the films. The SEM and AFM samples were prepared by spin-coating the graphene dispersions. **Figure 2b** shows that Med$_{Org}$ is composed of crumpled flakes with a nominal size of < 500 nm. The resulting flake network has a roughness of $S_a$ = 25 nm ± 5 nm. In contrast, the Low$_{Org}$ and High$_{H2O}$ dispersions both consisted of flat flakes with similar lateral sizes of < 1000 nm. A noteworthy difference between them is the measured flake network roughness of $S_a$ = 8 nm ± 5 nm for Low$_{Org}$ and $S_a$ = 3 nm ± 1 nm for High$_{H2O}$ (**Figure 2c**, **Figure 2d**). The difference can be attributed to the flake thicknesses ($t$), which were manually extracted from the AFM topography analysis as $t$ = 6,3 nm ± 2 nm for Low$_{Org}$ and $t$ = 4,6 nm ± 1 nm for High$_{H2O}$. Table 1 provides an overview of the relevant dispersion properties and the respective graphene flakes used in this work.

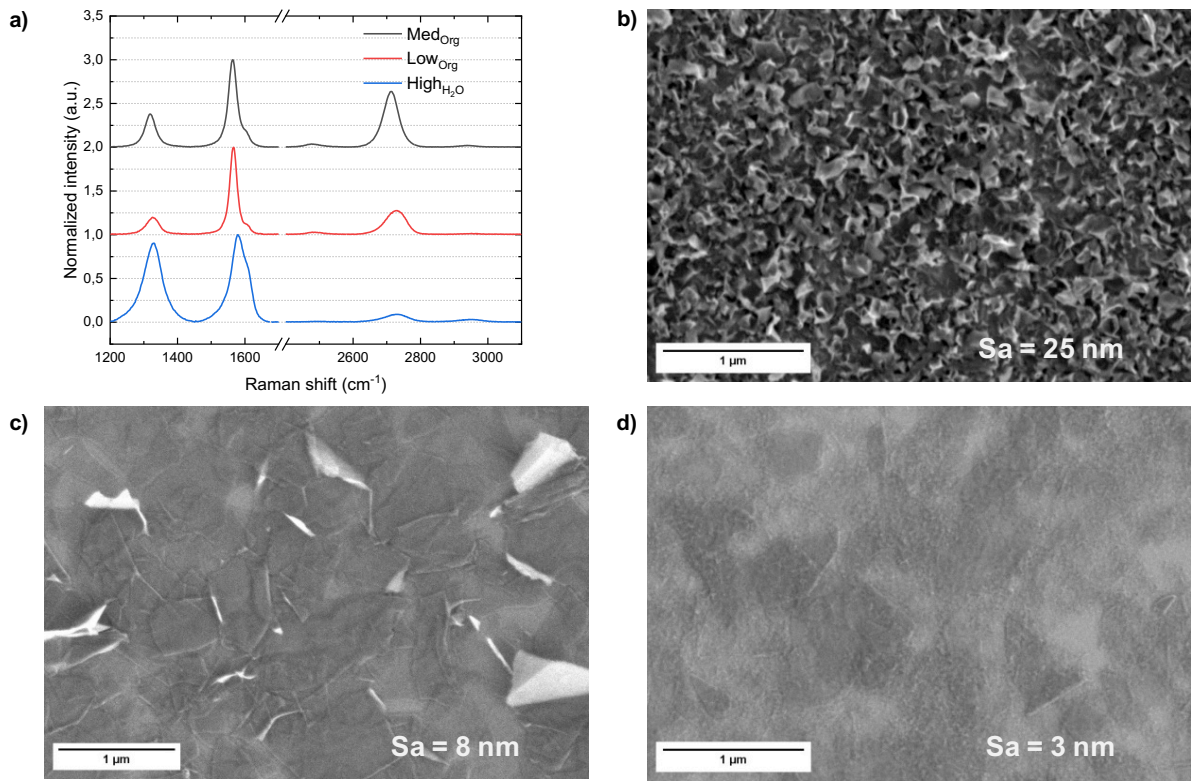

**Figure 2:** Characterization of the flake defect density via Raman spectroscopy and the flake network structure via SEM analysis and AFM roughness measurements (values in white). a) The comparison of the Raman spectra allows for the differentiation of graphene flakes on the basis of their defect density. We classify our graphene dispersions (and the corresponding samples) into medium (Med$_{Org}$), low (Low$_{Org}$), and high (High$_{H2O}$) defect densities. b) SEM micrograph and AFM roughness measurements of Med$_{Org}$ visualizing the crumpled graphene flake structure. c) SEM micrograph of Low$_{Org}$ showing flat, stacked flakes with relatively high roughness. d) SEM and AFM analyses confirm that High$_{H2O}$ consists of flat and thin flakes.

**Table 1**: Summary of the key characteristics of the graphene dispersions used and the resulting flake networks based on Raman, SEM and AFM analyses. The flake defect density is determined by the ratio between the D- and G-peak intensities extracted from the Raman

spectra. SEM enables its classification into crumpled and flat flake structures and an estimation of flake size. The surface roughness of the spin-coated flake networks was measured via AFM.

| Name | Flake defect density | Flake structure | Lateral flake size | Flake network roughness (Sa) |
|---|---|---|---|---|
| $Med_{Org}$ | $I_D/I_G = 0,38$ (medium) | Crumpled | < 500 nm | 25 nm ± 5 nm |
| $Low_{Org}$ | $I_D/I_G = 0,2$ (low) | Flat | < 1000 nm | 8 nm ± 5 nm |
| $High_{H2O}$ | $I_D/I_G = 0,9$ (high) | Flat | < 1000 nm | 3 nm ± 1 nm |

## 2.3 Shear strength measurements

We measured force–displacement diagrams for each sample (**Figure 3a)**. In each measurement, the initial segment of the curves, where the force $F = 0$ N, corresponds to the sample stage approaching the shear head. As mechanical contact is established, the force recording begins to increase until a maximum force is reached, indicating the shearing off of the button. After shearing, the force rapidly decreases to a value near zero again. The measured peak force represents the critical shear force $F_C$ of the button. The quotient of $F_C$ and the total button area of 100 x 60 μm² provides the shear strength $\tau_C$, our area-independent measure of the adhesion of the graphene flake network to the substrate.

Adhesion is a complex phenomenon that depends upon several factors, including intermolecular and interatomic interactions as well as surface roughness[55–57]. Furthermore, different methods for measuring adhesion might yield noncomparable results[58,59], making reproducibility challenging. Both facts underscore the importance of taking a statistically significant number of measurements per sample, which can increase reproducibility and reduce the effects of measurement variability. We performed BST and extracted $\tau_C$ on multiple buttons per sample. **Figure 3b** summarizes the results from this analysis, including the number of buttons per sample evaluated. For the PMMA$_{only}$ reference sample, $\tau_C$ = 33.37 ± 1 MPa was measured. First, this result validates the reproducibility of BST for adhesion measurements of 2D materials since the measured value is similar to the value obtained in different experiments (32.02 ± 1.7 MPa) reported by Schätz et al.[60]. Second, the $\tau_C$ of the PMMA$_{only}$ sample is at least two times greater than those of all the other investigated graphene dispersion samples. This significant reduction in $\tau_C$ in the presence of graphene under PMMA, as observed in all the samples, confirms that BST primarily probes the adhesion at the graphene-SiO$_2$ interface. This renders it suitable for assessing the adhesion of graphene flake networks on SiO$_2$.

The three dispersions contain flakes that differ in structure and defect density (see **Figure 2** and **Table 1**). The effects of these parameters on adhesion are investigated via a comparative analysis. Flake networks based on both planar flake types, Low$_{Org}$ and High$_{H2O}$, exhibit shear strengths of $\tau_C$ = 6.84 ± 0.7 MPa and $\tau_C$ = 14.6 ± 0.2 MPa, respectively. These values are higher than those for the crumpled flake type, Med$_{Org}$, with $\tau_C$ = 4.89 mN ± 0.8 mN. We attribute this to the increased planarity of both the flat graphene flake types and the resulting flake networks, which increases the contact area between the substrates and the bottom layers of the flake networks. Next, the impact of flake defect density on adhesion was evaluated by comparing both flat flake types (High$_{H2O}$ and Low$_{Org}$). High$_{H2O}$ results in a $\tau_C$ approximately two times greater than that of Low$_{Org}$, indicating a correlation between the defect density and shear strength. This finding is consistent with other works dealing with the adhesion of graphene materials[61–63]. According to these authors, defects, especially those capable of forming hydrogen bonds, favor adhesion. We compare our results with BST measurements of CVD monolayer graphene grown on copper and transferred to thermal SiO$_2$ by wet etching, as presented by Schätz et al.[51] (see the dashed line in **Figure 3b**). Our measurements demonstrate a higher $\tau_C$ for all the graphene flake types used in this work with respect to CVD graphene. These results further emphasize the correlation between defect density and adhesion and confirm the increased mechanical robustness of graphene flake networks with respect to CVD graphene.

HMDS is widely used in optical lithography as an adhesion promoter for photoresists. In the case of graphene dispersions, HMDS-coated substrates have been shown to prevent the coffee ring effect on inkjet-printed devices. However, in these works, the adhesion-promoting effect of HMDS on graphene flakes was only speculated[20,64]. We compared the $F_C$ measured on Med$_{Org}$ with that of the HMDS-coated Med$_{Org}$ sample (Med$_{OrgHMDS}$). The shear strength measured on Med$_{Org}$ is $\tau_C = 4.89 \pm 0.8$ MPa, whereas the value for Med$_{OrgHMDS}$ corresponds to $\tau_C = 7.97 \pm 1.8$ MPa. This finding primarily demonstrates the ability of HMDS to improve the adhesion of graphene flake networks on SiO$_2$.

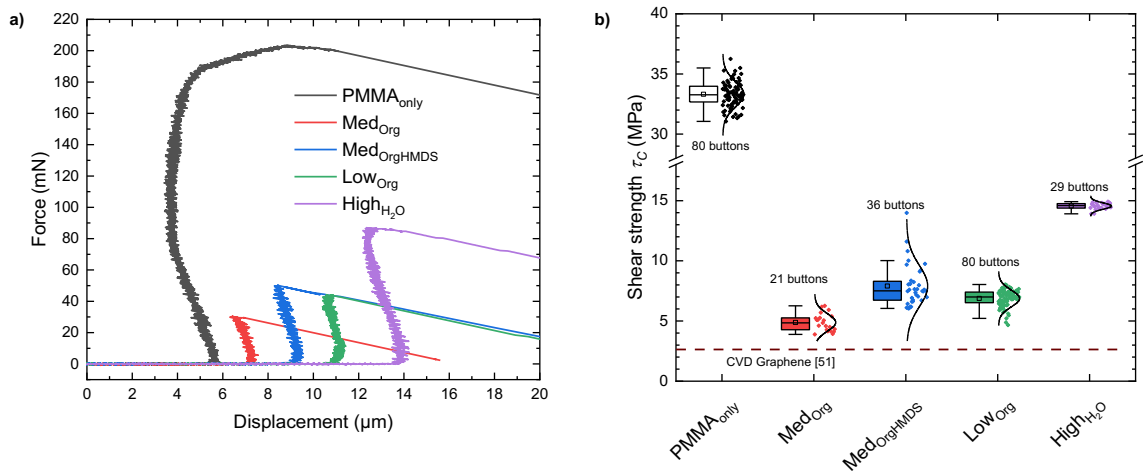

**Figure 3:** Adhesion measurements of different graphene flake types via the button shear test: extraction and evaluation of the critical shear force. a) Plot showing one selected force–displacement curve per sample. During the first phase with Force $F = 0$ N, the button approaches the shear head. Following the mechanical contact between the shear head and the button, the measured force increases until it reaches the critical shear force $F_C$. The x-axes of some force–displacement curves are shifted for better visualization. b) Box plot of $\tau_C$ showing comparisons between samples. All the graphene samples are far below the PMMA$_{only}$ sample, validating the method as an effective way to address the adhesion of graphene flakes. Flat flake types (High$_{H2O}$, Low$_{Org}$) have significantly greater delamination forces than do crumpled flakes (Med$_{Org}$), with the highest $\tau_C$ for High$_{H2O}$. HMDS significantly increases the measured critical shear force on crumpled flakes (Med$_{OrgHMDS}$).

**2.4 Flake type-specific degree of delamination**

We combined optical microscopy with Raman spectroscopy to assess flake type-specific delamination after BST. **Figure 4** shows one delaminated button per sample. The left parts of **Figure 4a**, **Figure 4b**, **Figure 4c**, and **Figure 4d** correspond to the micrographs of Med$_{Org}$, Med$_{OrgHMDS}$, High$_{Org}$ and Low$_{H2O}$, respectively. Three square regions (10 × 10 μm$^2$) were selected on each button. Each square is intended to reflect a different degree of delamination. Next, we split each image into its primary color channels via the software tool ImageJ[65], selected *red* and displayed it as a heatmap for better visualization (**Figure 4a**, **Figure 4b**, **Figure 4c**, **Figure 4d**, center). After the residuals from BTS were visualized, the clarification as to whether these are graphene flakes was provided by Raman analysis. We took Raman maps of the same fields as the heatmaps. The Raman maps represent the $I_G/I_{Si}$ ratio (**Figure 4a**, **Figure 4b**, **Figure 4c**, **Figure 4d**, right), which is proportional to the graphene thickness[66,67] and should best correlate with the laser microscopy analysis. We defined the degree of delamination ($\Delta_{Del}$) as the percentage ratio of the completely delaminated area to the total button area and quantified it on micrographs of four different buttons per sample via ImageJ.

The heatmaps and Raman maps in **Figure 4** show good correlations for all the samples and confirm that delamination after BST is incomplete (further analysis based on AFM and SEM is provided in the Supplementary Information). Additionally, the highest amount of residual graphene was found at the button edges. We attribute this to incidental O$_2$ plasma exposure of the PMMA sidewalls during button fabrication, which is not prevented by the overlying Al layer. This may alter the chemical composition and mechanical properties of PMMA[68,69] and likely reduce its adhesion to graphene. Ultimately, this may promote the delamination of PMMA from graphene rather than graphene from the substrate. All buttons show a consistently increased amount of residual graphene, especially at the left edge (shear direction from left to right). In this case, "mode mixing" should be considered an additional influencing factor. Mode mixing describes the relationship between the vertical tension force (Mode I) and the horizontal shear force (Mode II) applied to the button during the shear process[70]. It is characterized by variations along the shear direction, which are most pronounced in the initial stages of the shear process[49,71,72]. Therefore, significantly different delamination behavior can be expected at the shear edge of the button compared to the rest of the button[73].

**Figure *4*a** and **Figure *4*b** show that Med$_{OrgHMDS}$ has a greater degree of delamination ($\Delta_{Del}$ = 72.5 ± 5.7 %) than does Med$_{Org}$ ($\Delta_{Del}$ = 61 ± 5.2 %), despite the higher $\tau_C$ measured on Med$_{OrgHMDS}$ (see Figure 3). A similar inverse relation between $\Delta_{Del}$ and $\tau_C$ is observed for Low$_{Org}$ (Figure 4c) and High$_{H2O}$ (Figure 4d). Whereas High$_{H2O}$ shows almost complete delamination ($\Delta_{Del}$ = 96.6 ± 1.1 %), there is a significant amount of residual graphene over the entire button surface on Low$_{Org}$ ($\Delta_{Del}$ = 50.3 ± 8.8 %). We associate this with differences in interflake adhesion and propose evaluating it by analyzing the sample-specific degree of delamination. The reasons are twofold: First, our method probes the adhesion of the weakest material interface. Second, interflake junctions determine the mechanical properties of graphene flake networks[12,45]. Consequently, a well-conformed graphene flake network should result in stronger interflake than flake-substrate adhesion. This, in turn, should lead to delamination of the entire flake network. Conversely, poor interflake adhesion should promote partial delamination. We further attribute the greater degree of delamination of Med$_{OrgHMDS}$ than that of Med$_{Org}$ to enhanced interflake adhesion due to better flake network

formation in the presence of HMDS[20]. The improvement in network formation is based on the more densely packed flake network with greater flake overlap[14] and reduced interflake distances[74], as the HMDS-coated surface results in less spreading of the graphene dispersion[20,64,75]. This hypothesis is supported by our 4-point probe sheet resistance measurements (Supplementary Information), where the use of HMDS results in lower resistance values. The stronger interflake adhesion of $Low_{Org}$ compared to $High_{H2O}$ has two possible explanations, which are supported by our experiments: a strong hydrogen bond interaction due to its higher hydroxyl content[76–78], as evidenced by its higher defect density (see $I_D/I_G$ in Table 1), and better conformity due to its reduced flake thickness[12,38,75,79] (see roughness in Table 1). The parameters influencing the adhesion of the graphene flake networks are summarized in Table 2.

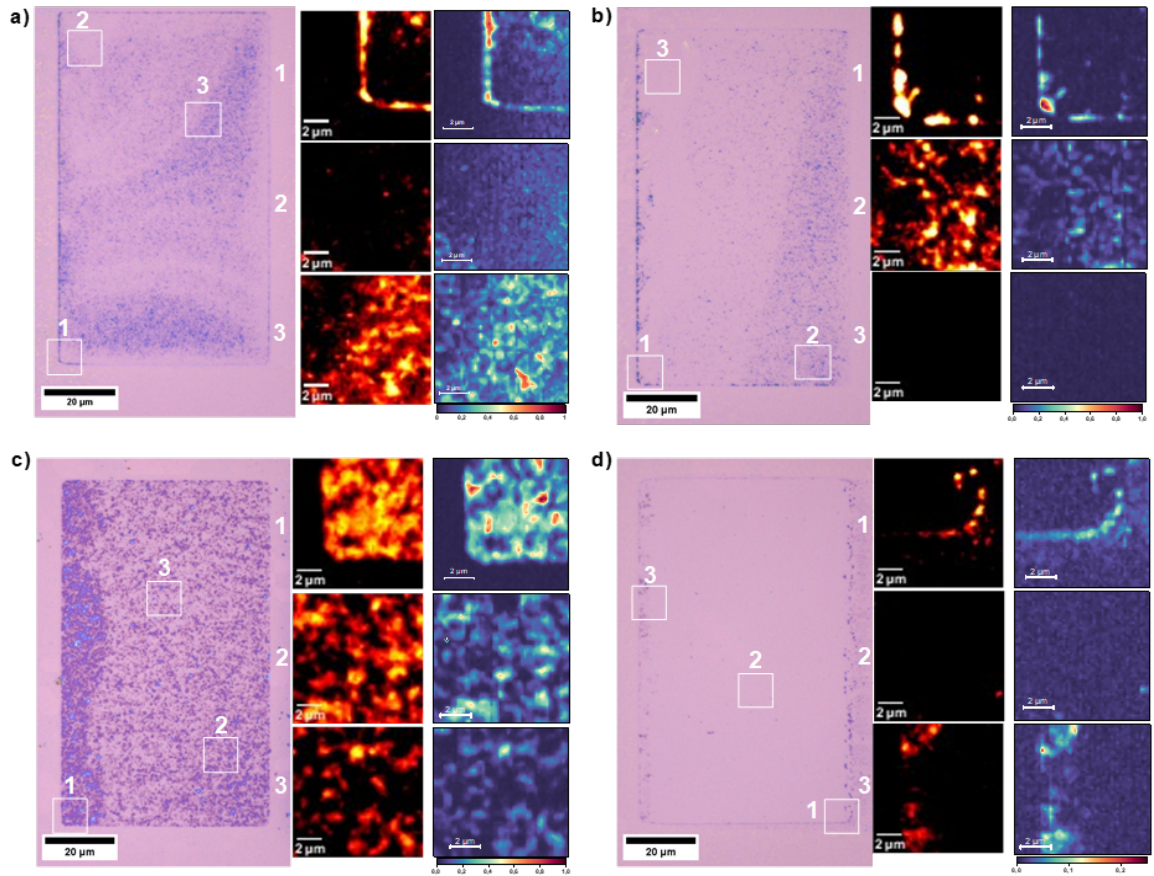

**Figure 4:** Optical and Raman analysis of the sample-specific degree of delamination. One subfigure per sample: a) Med$_{Org}$, b) Med$_{OrgHMDS}$, c) Low$_{Org}$, d) High$_{H2O}$. <u>Left in each subfigure</u>, the micrograph of the delaminated button (shear direction from left to right) includes three numbered fields (10 × 10 μm$^2$) at three distinct button regions. Each sample is representative of different degrees of delamination. <u>Center in each subfigure</u>: Magnified micrographs of marked sections, presented as heatmaps, provide a clearer visualization of residual material. <u>Right in each subfigure</u>: Raman maps of $I_G/I_{Si}$ taken at selected fields enable the confirmation of graphene residues and their correlation with thickness. The highest concentration of residual graphene flakes at the button edges, particularly at the shear edge, can be explained through the effect of O$_2$-plasma on PMMA and through mode mixing. The degree of delamination is related to interflake adhesion: the stronger the adhesion between flakes is, the more likely it is that the entire graphene flake network delaminates after BST.

**Table 2**: Summary of the assessed parameters and their influence on adhesion according to the BST results.

| Parameter | Effect on adhesion |
|---|---|
| Flake flatness | ↑ |
| Low defect density | ↓ |
| HMDS auf SiO$_2$ | ↑ |
| Strong interflake adhesion | ↑ |

## 3. Conclusions

We applied the button shear test method to assess the adhesion of graphene flake networks on SiO$_2$ and confirmed its efficiency, reliability, and repeatability as an adhesion measurement technique for graphene dispersion-based thin films. Our analysis included three different dispersions with different graphene flake types with respect to structure and defect density. All of them have higher delamination forces than CVD monolayer graphene does, confirming their increased mechanical robustness. Our work demonstrates that the flake structure and defect density are critical parameters for adhesion. We found that flat flake structures and their resulting networks lead to large contact areas, increasing adhesion. Furthermore, higher defect densities are beneficial for adhesion due to the increased presence of hydrogen bonds. Moreover, HMDS increases the adhesion of graphene flakes by improving film formation and consequently increasing interflake adhesion. Finally, we combined optical microscopy with Raman spectroscopy to analyze the delamination patterns resulting from the button shear tests. We found flake-type-specific degrees of delamination that can be correlated with interflake effects. It also reinforces the key role of the flake structure and defect density for both interflake and flake-to-substrate adhesion. Overall, our work establishes button shear testing as a quantitative technique for measuring the adhesion of graphene flake networks, more specifically, graphene inks dispersed on chip surfaces. We demonstrate the method with a comparative analysis that highlights material-specific differences in adhesion. This method, including our study, can be used to select suitable flake and dispersion types and their integration processes for specific target product applications. In addition, the consistent quantification of adhesion can be helpful in deriving mechanical robustness models aimed at product reliability.

## 4. Methods

*4.1 Button fabrication*

For all the samples, the substrate consisted of 280 nm thermally grown $SiO_2$-on-Si cut into $2 \times 8$ cm$^2$ wafer pieces.

*4.1.1   Reference sample (PMMA$_{only}$)*

Polymethyl methacrylate (PMMA) was spin-coated on the substrate to form a 5 µm thick layer. A 40 nm layer of aluminum (Al) was thermally evaporated on top. An approximately 800 nm thick layer of the positive photoresist S1805 from MicroChem was spin-coated and patterned to 100 x 60 µm$^2$ buttons via optical lithography. The tetramethylammonium hydroxide (TMAH)-based developer removes the photoresist and etches Al outside of the button area in the same step. Oxygen ($O_2$) plasma is performed via a MyPlas system from Plasma Electronics to remove PMMA and ultimately form buttons.

*4.1.2   Graphene dispersion samples without HMDS (High$_{H2O}$, Med$_{Org}$, Low$_{Org}$)*

The surfaces of the wafer pieces were treated with $O_2$ plasma. The graphene dispersion was deposited on the substrate by spin-coating at 1500 rpm. Successive spin-coatings were required to achieve full flake coverage of the substrate, which was verified by laser scanning micrographs. After dispersion deposition, the samples were thermally annealed for 30 minutes at 350 °C (Med$_{Org}$ and Low$_{Org}$) and 15 minutes at 250 °C (High$_{H2O}$) in an uncontrolled laboratory atmosphere. The subsequent button fabrication followed the steps described in **Section 4.1.1**.

*4.1.3  Graphene dispersion sample with HMDS (Med$_{OrgHMDS}$)*

HMDS was evaporated onto the substrate via a vacuum desiccator. Immediately after HDMS evaporation, Med$_{Org}$ is spin-coated and annealed as described in **Section 4.1.2.** The subsequent PMMA deposition and button patterning were performed according to **Section 4.1.1**.

*4.2  Button shear testing*

A Nordson Corporation DAGE4000Plus pull-shear tester was used to perform the button shear tests. The sample stage housed the sample containing the buttons. A 120 μm wide shear head was placed in front of the button and lifted approximately 2 μm above the sample surface. The measurement starts with the movement of the stage toward the shear head. The force acting on the shear head and the lateral displacement of the stage are then recorded. The sample stage velocity is set to 10 µm/s.

*4.3  Raman measurements*

The Raman measurements for dispersion characterization were performed on 280 nm thermal SiO$_2$-on-Si chips with drop cast graphene dispersions via a ProRaman-L Raman spectrometer. The Raman maps of the delaminated buttons were obtained via a WITec Alpha 300R Raman microscope. An excitation wavelength of 532 nm was selected for all Raman measurements regardless of the measurement setup.

*4.4  AFM measurements*

A Park NX20 atomic force microscope in noncontact mode was used for roughness measurements. The samples consisted of $SiO_2$-on-Si substrates spin-coated with each graphene dispersion and subsequently annealed as detailed in **Section 4.1.2**. The flake thickness was manually extracted via the software Gwyddion[80].

**Author Contributions**

M.C.L., U.K., and J.E.A.E. designed the study. A.W. and J.E.A.E. characterized the graphene dispersions/flakes. A.W., J.S., and J.E.A.E. fabricated the samples. J.S. and J.E.A.E. were involved in the button shear test measurements. J.R. conducted the Raman map measurements. J.E.A.E. performed the data analysis. M.C.L. J.E.A.E., J.S., J.R., D.T., and A.Z. discussed the data. The manuscript was written through the contributions of all the authors. All the authors approved the final version of the manuscript.

**ACKNOWLEDGMENT**

M.L. and J.R. acknowledge funding through the German Ministry of Economic Affairs and Climate Action (BMWK) through the project KATHOGRAPH (Grant ID: 01 1F23219N).

# Supplementary Information

# Measuring the Adhesion of Graphene Flakes via Button Shear Tests

*Adatti Estévez J. E., Schaetz J.[2,3], Ruhkopf J.[2,4], Weber A. F.[1], Tumpold D.[1], Zoepfl A.[3], Krumbein U.[1], Lemme M.C.[2,4]*

*Jorge Eduardo Adatti Estévez[1,2], Josef Schätz[2,3], Jasper Ruhkopf[2,4], Annika Weber[1], David Tumpold[1], Alexander Zöpfl[3], Ulrich Krumbein[1], Max Christian Lemme[2,4]\**

[1]Infineon Technologies AG, Am Campeon 1-15, 85579 Neubiberg, Germany

[2]Chair of Electronic Devices, RWTH Aachen University, Otto-Blumenthal-Str. 2, 52074 Aachen, Germany

[3]Infineon Technologies AG, Wernerwerkstraße 2, 93049 Regensburg, Germany

[4]AMO GmbH, Otto-Blumenthal-Str. 25, 52074 Aachen, Germany

*email: max.lemme@eld.rwth-aachen.de

**Analysis of graphene flake residuals by scanning electron microscopy (SEM)**

The use of SEM together with Raman analysis (Figure 4) allows for a comprehensive evaluation of the residual material following the Button shear test (BST) and confirms the nature of the detected residues as graphene flakes. In addition, SEM reveals graphene flakes in regions that, according to Raman and optical microscopy, appeared to be completely delaminated. This is the case for $Med_{Org}$ (Figure S1a) and $Med_{OrgHMDS}$ (Figure S1b), where single flakes are observable between larger flake clusters. Figure S1c shows the delamination pattern of $Low_{Org}$ with characteristic strong thickness variation of residual flakes. Figure S1d shows large residual flakes along the shear edge and large regions on $High_{H2O}$ that are residue free.

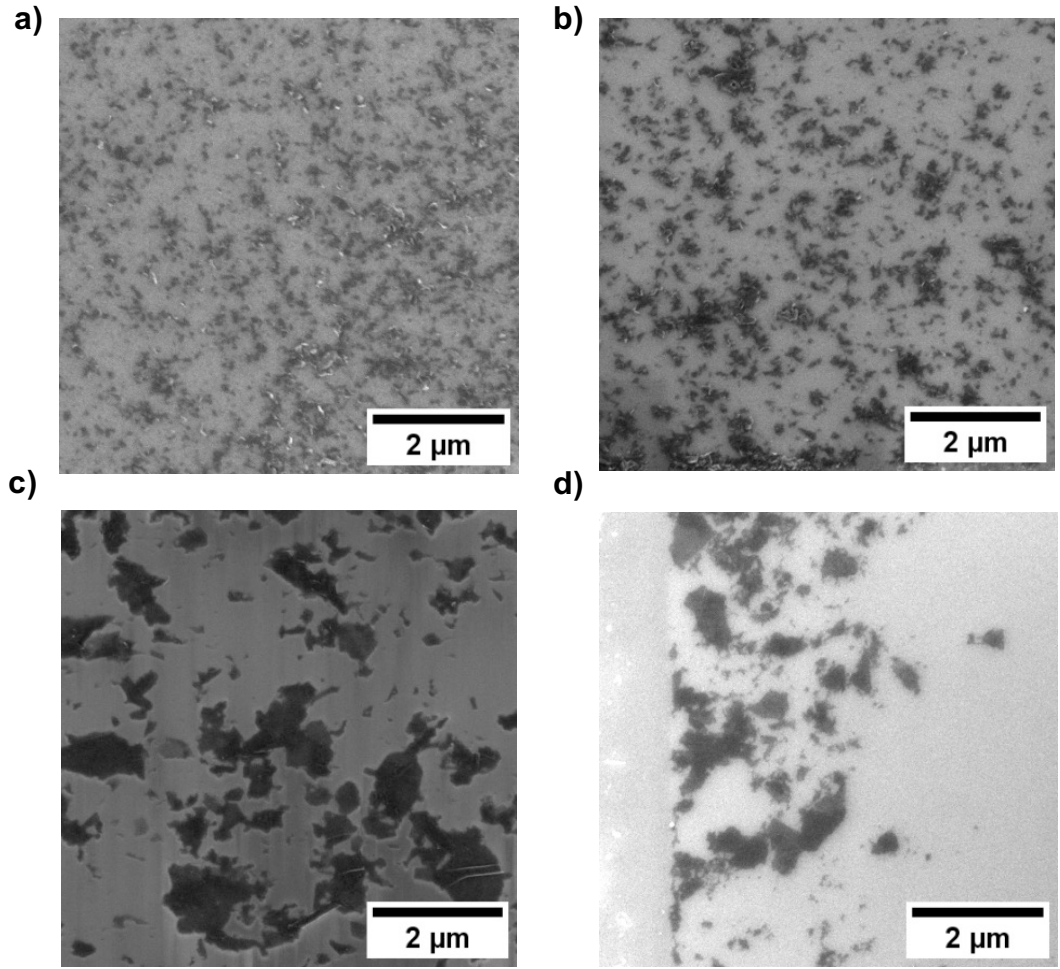

**Figure S1:** Observation of residual graphene flakes via SEM. a) SEM image of Med$_{Org}$ taken in Field 3 showing randomly distributed graphene flakes and areas of complete delamination. A noncontinuous flake network as a result of BST is visible. b) SEM image of Med$_{OrgHMDS}$ from Field 2 in Figure 4b: the graphene flakes are randomly distributed along the whole surface. Single graphene flakes are visible between larger flake bundles. c) Apart from the regions of total delamination, on Low$_{Org}$ (Field 3), the delamination pattern consists of flat flakes of different sizes and thicknesses. d) A significant amount of residual flakes is evident along the shear edge on High$_{H2O}$. The sizable residue-free areas are representative of most of the button area.

**Analysis of graphene flake residuals via atomic force microscopy (AFM)**

AFM complements Raman analysis in determining if the residues observed are graphene flakes (Figure S2). The presence of graphene residues was confirmed in all four samples. In addition, AFM revealed graphene flakes in regions that appeared to be completely delaminated. This is the case for $Med_{Org}$ (Figure S1a) and $Med_{OrgHMDS}$ (Figure S1b). For $Low_{Org}$ (Figure S1c), a significant number of flakes can be observed. Single flakes, flake stacks of different thicknesses, and regions free of flakes are visible. This further emphasizes the partial delamination noted in the Results section. Figure S1d shows an extremely low density of residual flakes on $High_{H2O}$ after BST, which is in line with the discussion in the Results section.

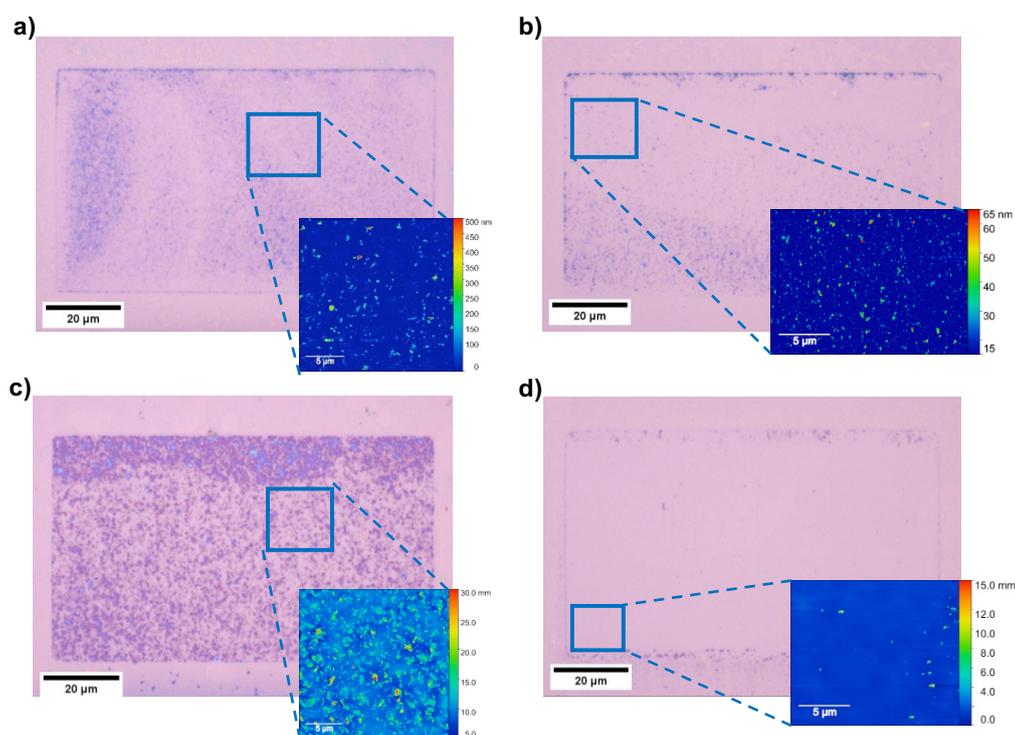

**Figure S2**: AFM analysis confirming the presence of graphene flake residues on all analyzed samples after BST. a) $Med_{Org}$ and b) $Med_{OrgHMDS}$ show a similar behavior of randomly distributed single flakes. c) Single flakes and flake stacks of different thicknesses are visible in $Low_{Org}$. The resulting flake network lacks continuity, as can be observed by the regions free of flakes. d) The low allocation of single flake or flake stacks on $High_{H2O}$ indicates almost complete delamination.

**Sheet resistance measurements on Med$_{Org}$ and Med$_{OrgHMDS}$**

The four-point probe (4PP) method, implemented with a Jandel RM3-AR system, enables straightforward and efficient measurement of the sheet resistance of as-deposited graphene flakes. We measured the sheet resistance of two Med$_{Org}$ and two Med$_{OrgHMDS}$ samples. The samples were fabricated following the same procedure as the BST samples (see Experimental Section), except for the PMMA and Al deposition and structuring. The results are presented in Figure S3. The reduced sheet resistance in the presence of HMDS is attributed to a lower degree of spreading of the graphene dispersion. This results in shorter interflake distances, which in turn produce lower interflake resistances and ultimately a reduced sheet resistance[1,2].

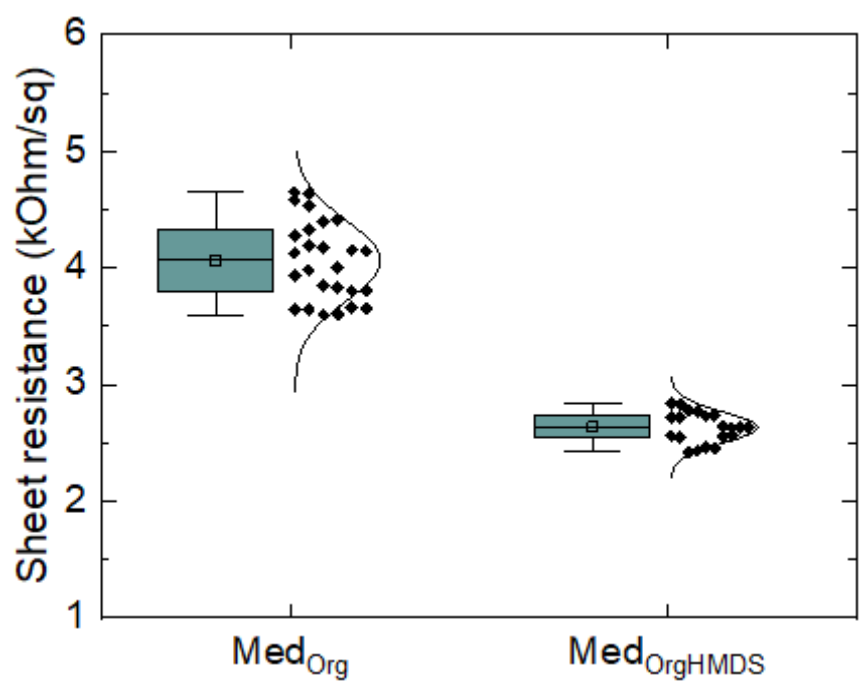

**Figure S3:** Evaluation of the effect of HMDS on the electrical resistance of graphene films based on crumpled flakes. The Med$_{Org}$ samples consistently show higher sheet resistances than the Med$_{OrgHMDS}$ samples despite having identical spin–coat parameters. This is due to HMDS, which increases the contact angle and leads to low spreading of the graphene dispersion, resulting in shorter interflake distances and lower interflake resistances.